# Being Together in Place as a Catalyst for Scientific Advance

Eamon Duede[1,2,8*]

Misha Teplitskiy[3,6]

Karim Lakhani[4,5,6]

James Evans[7,8,9*]

[1]Digital, Data, and Design Institute, Harvard University, Cambridge, MA, 02138

[2]Embedded EthiCS, Harvard University, Cambridge, MA, 02138

[3]University of Michigan School of Information, Ann Arbor, MI, 48104

[4]Harvard University, Cambridge, MA, 02138

[5]Institute for Quantitative Social Science, Harvard University, Cambridge, MA, 02138

[6]Laboratory for Innovation Science at Harvard, Boston, MA, 02134

[7]Sociology, University of Chicago, Chicago, IL, 60607

[8]Knowledge Lab, University of Chicago, Chicago, IL, 60607

[9]Santa Fe Institute, Santa Fe, NM 87501

**Abstract**:

The COVID-19 pandemic necessitated social distancing at every level of society, including universities and research institutes, raising essential questions concerning the continuing importance of physical proximity for scientific and scholarly advance. Using customized author surveys about the intellectual influence of referenced work on scientists' own papers, combined with precise measures of geographical and semantic distance between focal and referenced works, we find that being at the same institution is strongly associated with intellectual influence on scientists' and scholars' published work. However, this influence increases with intellectual distance: the more different the referenced work done by colleagues at one's institution, the more influential it is on one's own. Universities worldwide constitute places where people doing very different work engage in sustained interactions through departments, committees, seminars, and communities. These interactions come to uniquely influence their published research, suggesting the need to replace rather than displace diverse engagements for sustainable advance.

[*]Correspondence to: Eamon Duede, eduede@g.harvard.edu and James A. Evans, jevans@uchicago.edu

**1. Introduction:**

The COVID-19 pandemic necessitated social distancing at every level of society (Venkatesh and Edirappuli 2020), raising essential questions concerning the importance of place and proximity. Universities substituted face-to-face instruction, mentorship, faculty meetings, and research seminars with video conferences, not replacing but displacing interactions that otherwise lead to spill-over conversations and unintentional connections, which, in turn, spark innovative scientific and scholarly ideas and collaboration. With university laboratories only partially staffed, seminar rooms empty, workshops closed to outsiders, and conferences made either hybrid or virtual, delayed, or indefinitely rain checked, questions arise regarding the effect that this social distancing will have on scientists' knowledge of emerging ideas and findings, and their ability to influence and be influenced by one another on the path to discovery and collective advance (Akour et al. 2020). With many organizations still operating in a mostly 'hybrid' or fully remote manner, while others are in the process of returning or have fully returned to the office, unique, natural experiments are currently underway that will add critical insight to these questions (Rosen 2021; Brucks and Levav 2022).

Researchers have examined the effect of geographical distance on the practice of both science and invention. Ubiquitous digitization, virtual classrooms, workshops, and conferences have led some to declare a "death of distance" not only in the world but also in science (Cairncross 1997; Friedman 2006). Recent experience and employee interest have led many businesses and institutions to announce that they plan to make remote work a more permanent feature of their organizational structure. But does the collapse of distance with advances in transportation and communication technology remove the geographic agglomeration that has always characterized the production and consumption of complex scientific and technical knowledge (Collins 1974; Jaffe, Trajtenberg, and Henderson 1993; Evans 2010; Murata et al. 2014)? Recent high-profile commentary argues that there is no support for creative contributions catalyzed by being together in place (Miller 2021) or that findings are mixed (Mors and Waguespack 2021). Yet new research implies that geography may still matter.

Distance has been found to be a significant factor in conditioning collaboration (Olechnicka, Ploszaj, and Celińska-Janowicz 2018; Morgan 2004; Olson and Olson 2000; Adams 2013; Fernández, Ferrándiz, and León 2016; Catalini 2018; Criscuolo and Verspagen 2008). In a recent study involving tens of thousands of information workers at a major technology company, remote work during COVID-19 shutdowns resulted in a more siloed, static, and asynchronous collaboration network, with fewer bridges between disparate parts of the firm network to facilitate rapid information flow (Yang et al. 2021). These findings suggest that geographical co-location is an important factor in linking individuals whose immediate intellectual agendas are relatively distant, and motivates our investigation into the interaction of geographical and intellectual distances. While there is some evidence that regional scales affect knowledge spillovers with scholarly citation decreasing with distance (Wichmann Matthiessen, Winkel Schwarz, and Find 2002; Börner et al. 2006), most research on distance has tended to focus on inventive activity where patent citing practices remain distinct from those in science (Criscuolo and Verspagen 2008). All of this work suggests that distance matters for increasing awareness of relevant research, but these studies also suggest that what it means to be "close" can be as far as the same country, region, or within hundreds of miles. Not very close. If this diffuse geographic influence were all to the story, then the dissipation of concrete university settings into clouds would be less cause for concern.

Attempts have been made to "zoom in" on micro, hyper-local scales, with a focus on particular institutions and even buildings. While macro distances seem to play an important role in determining what prior knowledge scientists are more likely to cite, micro distances have been shown to reorient research directions and productivity (Rawlings and McFarland 2011), catalyze the consummation of research relationships (Kabo et al. 2015a), facilitate the transfer of skills and tacit knowledge (Collins 1974; Collins and Harrison 1975), and promote the consolidation of distinct and diverse epistemic cultures (Cetina 2009). From this work, we learn that researchers are more likely to be productive when they learn and adopt best practices from their peers, that shared interactions are more likely to lead to proposed innovations, and that universities can encourage this by promoting collaboration among researchers (Rawlings and McFarland 2011; Kabo et al. 2015b). In a more recent study, researchers found that co-location within an institution has the most substantial distance-related effect on the probability of citation, that sharing an institution improves the probability that intellectually distant works will be cited, and that research practice is learned informally, through interactions with advisors (Wuestman, Hoekman, and Frenken 2019; Leahey 2006).

Yet, it remains unclear whether the physically proximate, nearby work we are more likely to cite is vital for influencing our own work, or simply a curious but ornamental allusion. Moreover, while paths of possible collaborators are likely to cross more often in buildings and on campuses, how do the partnerships that emerge from those face-to-face interactions rank in importance and influence relative to those formed online and across longer distances?

Estimates of the probability of citation tell us something about the ways in which institutions facilitate access to scholarship, but they do not tell us enough about the impact of co-location on intellectual influence to shape policy. Institutions are faced with a costly decision for which current literature gives them little guidance. Does the probability of citation under- or overestimate the probability of influence? Depending on the answer to this question, a cost-benefit analysis may recommend enhanced virtualization or a wholesale return to physical co-location. As we show in this article, misspecifying the relationship between citation and influence could have substantial effects on innovative capacity moving forward.

The limitation of prior work in this area is that it focuses on how distance affects the probability of citation or collaboration, but leaves us in the dark about what it means for intellectual influence. Not all references denote intellectual impact (Bornmann and Daniel 2008; MacRoberts and MacRoberts 1996; Nicolaisen 2007). With some citations indicating meaningful influence and others denoting obligatory signals of membership within an intellectual community or the flex of intellectual control (Teplitskiy et al. 2022), attempts to observe the effect of distance on scholarship that rely primarily on the probability of citation cannot disentangle influence from exposure.

Moreover, observational and experimental studies of face-to-face versus virtual interaction suggest potential mechanisms for the difference between local, regional, or global engagement and collaboration. Virtual interactions limit the flow of subtle but important influences in online settings. Face-to-face interactions have been shown to amplify the impact of collective motivation on team performance (Kirkman et al. 2004). Direct experimentation with the use of video-conferencing in natural and controlled settings demonstrates how virtual interaction curbs collective creativity compared with face-to-face collaboration by focusing participants on a single channel of interaction—the screen (Brucks and Levav 2022).[1] While no controlled studies have examined the effect of virtual interactions on sustained intellectual influence at scale, existing studies suggest the importance of examining the relative effects of interactions over distances that would facilitate face-to-face versus computer-mediated interaction.

[1] Scholarship by boosters of communication technologies argue that there are few differences *in theory* between what virtual and face-to-face communication can achieve (Rhoads 2010), but these speculations are at variance with the empirical effects (e.g., Brucks and Levav 2022).

Our research design seeks to directly assess intellectual influence and knowledge transmission using publication data from Clarivate's *Web of Science* and surveyed author reports as in (Teplitskiy et al. 2022). Here, we initially summarize the process, then detail each step in the Methods section below. We began by randomly sampling seed articles from 15 diverse fields drawn from the physical sciences, life sciences, social sciences, and humanities. For each field, we randomly selected focal articles that cited these seed papers. We selected two references from each focal paper and surveyed corresponding authors of the focal papers regarding how much each referenced paper influenced the author of the focal paper, how well they knew it, as well as how and where they first discovered it (e.g., database, colleague, presentation). This yielded measurements of the intellectual influence, knowledge, and provenance of two referenced works for a total of 12,008 works (with some works rated by multiple respondents).

To understand how intellectual influence and knowledge transmission are related to physical proximity, we gathered information on the organizational and geographic locations of the home institutions for focal and referenced papers. To understand how intellectual influence is related to intellectual distance, we measured the intellectual (scientific and semantic) distance between referenced and focal papers with semantic precision by encoding a rich trace of the content (e.g. title and abstract) in a geometric embedding space using one of the unsupervised machine learning models that have come to exhibit human-level sensitivity in natural language tasks (Mikolov et al. 2013; Pennington, Socher, and Manning 2014; Peters et al. 2018; Devlin et al. 2018). Our analysis then involved regressing intellectual influence and knowledge on organizational and intellectual distance measurements to identify the relationship between organizational proximity and influence.

## 2. Methods

### 2.1 Sampling

We used data from Clarivate's complete *Web of Science* (WoS) database to systematically sample the scholarly literature and survey the scientific community across the following 15 fields indexed by WoS: biochemistry & molecular biology, physical chemistry, economics, endocrinology & metabolism, energy & fuels, electrical & electronic engineering, history & philosophy of science, immunology, linguistics, nanoscience & nanotechnology, oncology, pharmacology & pharmacy, applied physics, psychology, and telecommunications. Fields were chosen to provide broad disciplinary coverage and make our results as generalizable as possible. We selected the fields in the following way. WoS attributes journals to fields that fall under six major subjects - Arts & Humanities, Clinical, Pre-clinical & Health, Engineering & Technology, Life Sciences, Physical Sciences, and Social Sciences. We selected fields with large coverage by WoS from each of these major subject areas. We further selected on the extent to which citation-based metrics were meaningful for these fields by averaging the CiteScore (Elsevier's journal impact metric) for each field's top five journals in 2016, then ranking fields according to this average. Together, this ensured broad topical coverage with substantial citation attention.

Data were collected in 2018 via a personalized Qualtrics survey to randomly sampled corresponding authors of papers published in 2015 in the WoS database. The year 2015 was chosen because, when we initially designed the study, it was the most recent year of data in our version of the database. For each of 15 focal fields, we identified all research articles published in 2000, 2005, and 2010 (to provide substantial temporal variation) and, for each year separately, ranked them according to the number of citations they had accrued through 2015. From each percentile of the resulting year-specific citation distribution, we randomly selected five referenced papers. We then identified all papers citing these cited papers in 2015, and from this list of "citing" papers selected five at random. If a cited paper did not have five citing papers in our database in 2015, we selected another paper from the same citation percentile and repeated the procedure until we accrued 25 citing papers for that percentile for that year. In 2018, we contacted the corresponding author of each focal paper with a personalized survey and asked them about two references in this paper. We sought two references from each citing paper to enable analyses with author fixed-effects. If a second reference (cited paper) matching the criteria was not available, we relaxed the matching constraints until a reference could be found. (Teplitskiy et al. 2022) provides additional details on the sampling and survey.

### 2.2 Data on Intellectual Influence and Knowledge

Our survey focuses on identifying our two core dependent variables: how much a referenced paper influenced the citing author in writing their focal paper, and how well they know it ('influence' and 'knowledge', respectively). We measured how influential a referenced paper was to the focal paper with the question, "How much did this reference influence the research choices in your paper?" Answer choices ranged from 1 (very minor influence: paper would have been very similar without this reference) to 5 (very major influence: motivated the entire project). We measured how well the respondent knew the content of the referenced paper with the question, "How well do you know this paper?" Answer choices ranged from 1 (not well: only familiar with main findings) to 5 (extremely well: know it as well as my own work). This approach yields a direct measure of the intellectual influence that a particular referenced work had on an author's own work, and how well authors know the referenced papers they cited in their own work. These two variables serve as central dependent variables in our analysis (see Table 1 for descriptive statistics of these variables). The response rate for the personalized survey was measured by clicks on the personalized survey link. The rate varied significantly among different disciplines, with oncology having the lowest rate at 12.9% and history and philosophy of science having the highest rate at 34.1%. The response rate for this survey was significantly higher than those obtained by other recent (Myers et al. 2020; Radicchi, Weissman, and Bollen 2017) email-based surveys of researchers, ranging from 50% to 1000% larger. The number of completed surveys varied by field, yielding 1060 responses from biochemistry & molecular biology, 1361 from physical chemistry, 1078 from economics, 589 from endocrinology & metabolism, 1419 from energy & fuels, 688 from electrical & electronic engineering, 209 from history & philosophy of science, 622 from immunology, 421 from linguistics, 497 from nanoscience & nanotechnology, 701 from oncology, 834 from pharmacology & pharmacy, 864 from applied physics, 1096 from psychology, and 569 from telecommunications.[2]

### 2.3 Data on Geographical Proximity

In order to analyze how intellectual influence relates to geographical proximity, we gathered information on the geographic location of host institutions for focal and referenced papers.[3] To accomplish this, we first extracted institutional addresses for each paper's corresponding author recorded in the *Web of Science*. Next, we geocoded these addresses using the Google Maps API to resolve their precise latitude and longitude, as well as city, country, and institutional information.

---

[2] For a comprehensive non-response analysis of the data used in this study, see (Teplitskiy et al. 2022).

[3] In geographical studies, 'institutional distance' and 'proximity' often denotes differences in national institutions or varying institutional backgrounds, such as in university-industry partnerships. In this paper, are focus is primarily on universities which are commonly referred to as "institutions". Throughout, where ambiguity might arise, we adopt the locution "organizational proximity", but otherwise use the terms "institution" and "organization" interchangeably to refer primarily to institutions of higher learning: universities.

Additionally, we extracted the institution and department names from the WoS database and cross-checked the institution with the Maps API. Finally, we calculated the shortest geodesic distance between the institutions of each focal paper and its corresponding referenced paper in kilometers.

When evaluating the effect of continuous distance on intellectual influence, we found that effects were prominent when distances between focal and referenced papers were near zero. Figure 1 reveals that these effects are highly nonlinear, with local effects washed out by long distances. To model these organizationally relevant distances, we discretized geographical proximity into five categories: "same academic department," "same institution," "same city," and "same country." Descriptive statistics for data on physical distance are also presented in Table 1.

Figure 1: Scatter plot of influence and knowledge as a function of continuous geographical proximity. The blue and red curves are lowess curves for influence and knowledge, respectively.

## 2.3 Data on Reference Discovery

Additionally, we gathered information on how respondents found their referenced papers. Survey respondents were given the following options to indicate the pathway by which they found the referenced paper: 'I know the author personally', 'The reference was recommended by a colleague', 'I found the reference via a presentation', 'I found the reference via another paper', 'I found the reference via database search', 'Not sure', and 'Other'. We present descriptive statistics on pathways by which respondents discovered referenced papers in Table 1.

### 2.4 Data on Intellectual Distance

In order to analyze how intellectual influence relates to intellectual distance, we measured the intellectual distance between referenced and focal papers with semantic precision. To do this, we encoded a rich trace of article content (title and abstract) in a word embedding model using one of the unsupervised machine learning approaches that have transformed modern natural language processing (Mikolov et al. 2013; Pennington, Socher, and Manning 2014; Peters et al. 2018; Devlin et al. 2018). Word embedding models draw on large-scale text corpora and "discover" semantics from local linguistic context, validating the distributional hypothesis that words occurring in the same contexts tend to have similar meanings (Harris 1954) by performing at human-level on analogy tests (Mikolov et al. 2013; Pennington, Socher, and Manning 2014), question answering (Peters et al. 2018; Devlin et al. 2018), and a wide range of language understanding tasks. It has been demonstrated that embedding texts produced by persons in given times and places can replicate surveyed associations among people from those same times and places (Kozlowski, Taddy, and Evans 2019; Caliskan, Bryson, and Narayanan 2017; Lewis and Lupyan 2019; Garg et al. 2018). Here, we use a popular word and document embedding algorithm, the Gensim implementation of Doc2Vec (Le and Mikolov 2014), and calculate the inverse of intellectual distance as the cosine similarity between fixed-length feature vector representations encoding each referenced and focal paper. Cosine similarity is calculated as

$$\text{similarity}(A, B) = \frac{A \cdot B}{\|A\| x \|B\|} = \frac{\sum_{i=1}^{n} A_i \times B_i}{\sqrt{\sum_{i=1}^{n} A_i^2} \times \sqrt{\sum_{i=1}^{n} B_i^2}}$$

where vectors $A$ and $B$ are focal and reference document vectors, and values closer to 1 represent more similar documents. This approach produces estimates of greater semantic similarity than bibliometric approaches for assessing the co-citation of articles or journals (Hamers and Others 1989), while not assuming that compared works frame themselves with respect to the same prior work.

All papers in our evaluation sample are ones our respondents cited. However, in order to render the semantic space in such a way as to capture the intellectual search space respondents might have used when selecting papers to read and cite, we also gathered a 10% sample of all papers published in each of the years between 2010 and 2015. From this pool, we drew a 10% random sample with replacement. As a result, our embedding space contains the abstracts of 543,936 documents. Each vector in the space represents a document's concatenated 'title + abstract'. We used the following hyperparameters for training the Doc2Vec model: *min_count = 10, dm = 0, dbow_words = 0, window = 10, sample = 0.000001, negative = 5, vector_size = 300*

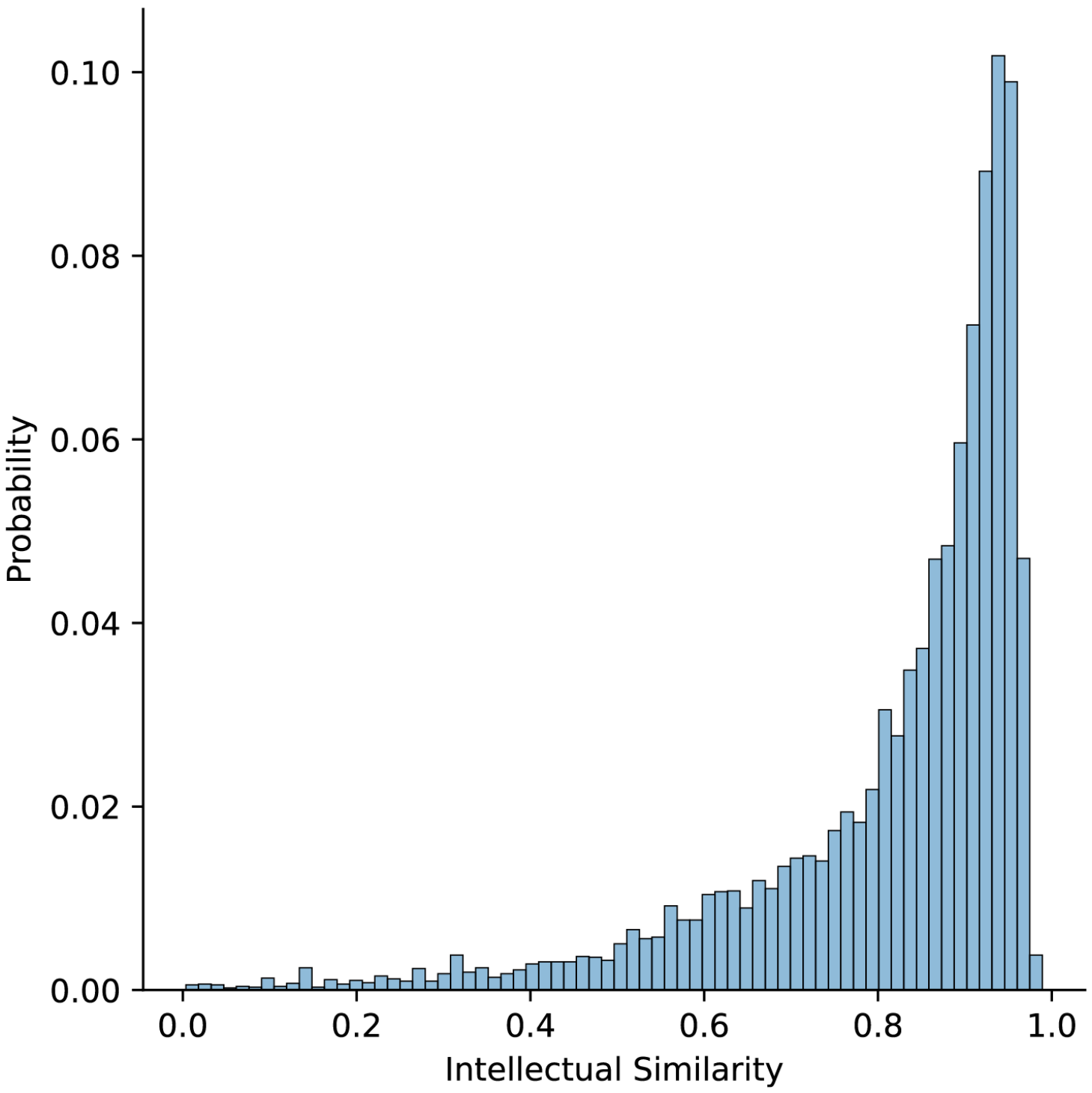

Figure 2: Probability distribution of cosine similarities between focal and reference paper pairs in our sample.

Similarity between pairs of focal and referenced documents in our corpus are right skewed (Figure 2), meaning that most referenced papers are very similar to the papers in which they are cited, as is to be expected. While we ultimately find that more intellectually distant works confer greater influence, here we observe the obvious: one cannot cite random papers. As a result, seemingly small movements in the direction of similarity or dissimilarity result in meaningfully large deviations from mean influence. As an added validation of the reliability of our embedding, we hand evaluated the 20 most similar and 20 least similar document pairs to ensure that comparisons were reasonable. We present descriptive statistics on intellectual distances between focal and referenced papers in Table 1.

Table 1: Descriptions and Summary Statistics of Variables

| Variable | Count | Mean | std | Source | Definition |
|---|---|---|---|---|---|
| influence | 12309 | 2.48 | 1.05 | survey | Survey response value for how influential the focal paper was on the citing paper (1-5) |
| knowledge | 12285 | 2.93 | 1.09 | survey | Survey response value for how well the respondant knows the content of the focal paper (1 - 5) |
| geo_distance | 10649 | 5787.97 | 4674.98 | google maps | Calculated geodesic distance between the focal paper's institution and citing paper's institution |
| intellectual distance | 12309 | 0.82 | 0.17 | language model | Cosine similarity between focal and citing paper abstract in vector space representation |
| same department | 12309 | 0.15 | 0.36 | web of science | Indicator variable for whether focal and citing paper share an academic deptartment |
| same institution | 12309 | 0.02 | 0.14 | web of science | Indicator variable for whether focal and citing paper share an institution |
| same city | 12309 | 0.01 | 0.11 | google maps | Indicator variable for whether focal and citing paper share a city |
| same country | 12309 | 0.16 | 0.42 | google maps | Indicator variable for whether focal and citing paper share a country |
| know personally | 12149 | 0.14 | 0.35 | survey | Survey boolean response variable indicating whether the respondant knows a focal paper author personally |
| colleague | 12149 | 0.10 | 0.30 | survey | Survey boolean response variable indicating whether the respondant learned of the focal paper from a colleague |
| presentation | 12149 | 0.05 | 0.23 | survey | Survey boolean response variable indicating whether the respondant learned of the focal paper from a presentation |
| another paper | 12149 | 0.25 | 0.43 | survey | Survey boolean response variable indicating whether the respondant learned of the focal paper from another paper |
| database search | 12149 | 0.64 | 0.48 | survey | Survey boolean response variable indicating whether the respondant learned of the focal paper from a database |
| not sure | 12149 | 0.04 | 0.20 | survey | Survey boolean response variable indicating that the respondant is unsure of how they learned of the focal paper |
| other | 12149 | 0.07 | 0.25 | survey | Survey boolean response variable indicating whether the respondant learned of the focal paper from some other source |

We represent the correlations between all variables used in the study in Appendix C, Table C2, where we observe that being at the same institution, influence, and knowledge of the referenced paper are each most positively correlated with 1) knowing the author personally, 2) having the author as a colleague, and 3) learning of the paper through a presentation or seminar. By contrast, being at the same institution, influence and knowledge of the paper are most negatively correlated with finding the referenced paper through a scholarly database, another paper, or not remembering how the paper was found.

In Figure 3, we further graphically explore the complex relationship between intellectual distance and the same or different institution. In that figure the red hand of each "clock" is fixed and represents the intellectual influence (thickness) and intellectual distance (length) of papers cited from other institutions relative to the focal paper at center. The blue hand represents the intellectual influence and distance of papers from the same institution, and the angle between red and blue is the intellectual distance between cited papers from the same and other institutions ($90^{o}$ indicates no semantic relation). For every field studied, articles cited from the same institutions are more distant and more influential. The inset above each clock face is a 2-dimensional UMAP projection (McInnes, Healy, and Melville 2018) of the position of focal papers relative to the papers they cite from other institutions (ends of the red line) and relative to those from the same institution (ends of the blue line.) Numbers orthogonal to each blue line are percentage increases in the intellectual influence of papers from the same relative to other institutions, and numbers along the blue line are percentage increases in the intellectual distance of papers from the same relative to other institutions in the uncompressed 300-dimensional semantic space in which they were embedded. Again, we clearly see that in every field studied, same institution is both associated with greater intellectual distance and influence.

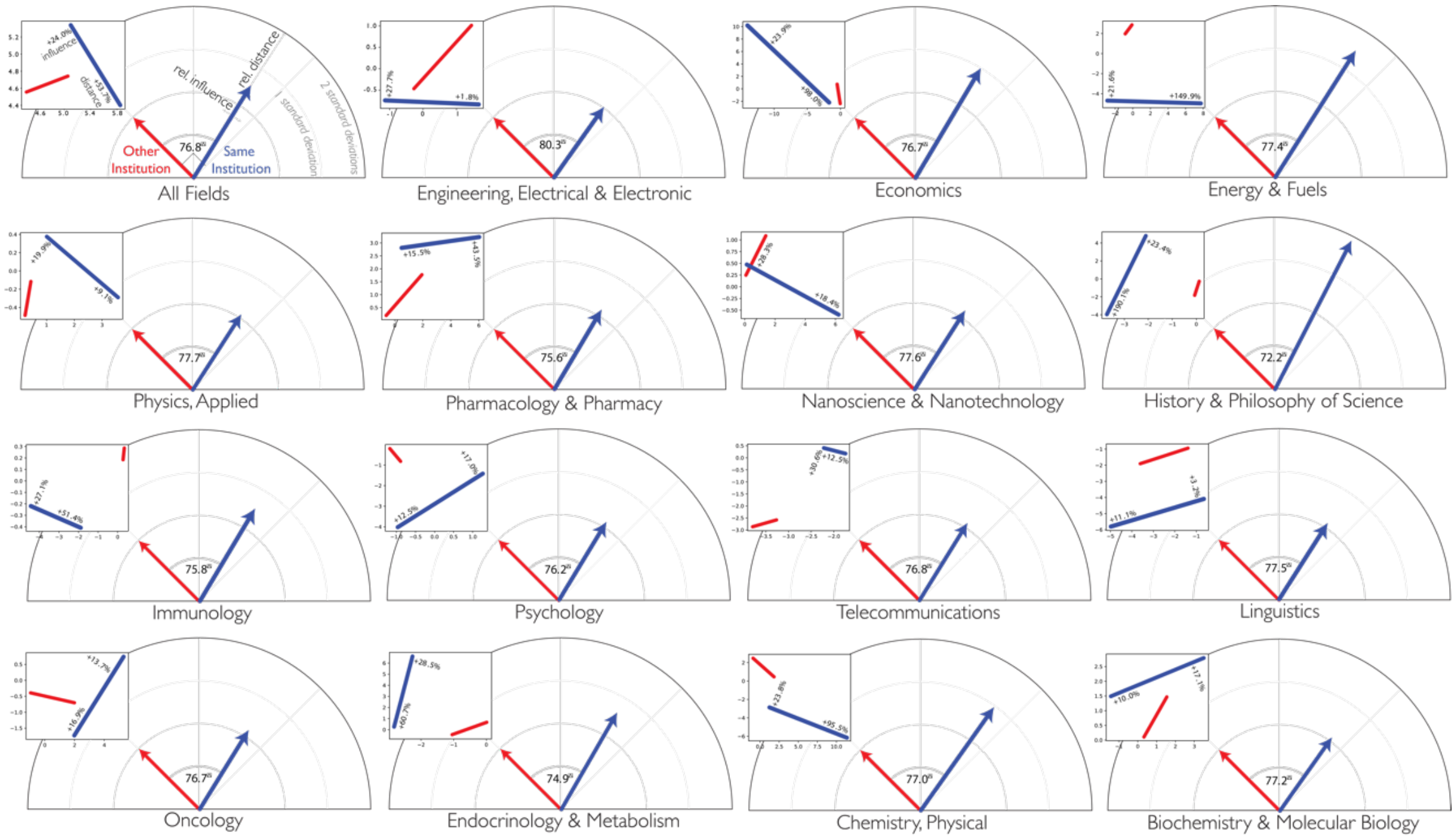

Figure 3: Influence and Intellectual Distance of Papers from same versus other institutions.

These and other associations presage many of the findings in our analyses below, but because correlations cannot enable author and paper fixed effects, our analyses below offer far greater quasi-experimental control.

### 2.5 Modeling the Structure of Influence

We initially evaluate the linear impact of our discrete measures of geographical proximity on influence and knowledge with the following specification:

$$y_{ij} = \alpha_i + \beta_0 + \beta \mathbf{X}_{ij} + \varepsilon_{ij}$$

where (as above) $y$ represents influence or knowledge, and the vector of independent indicator variables $\mathbf{X}$ represent the discrete distance measures (e.g., same department, same institution, same city, same country, and same world). Indices $i$ and $j$ enumerate authors and referenced papers, respectively. Author-fixed effects $\alpha_i$ denote author-specific intercepts and allow us to control for all (stable) differences between authors, including their fields. The results of this regression (Table 2 and Figure 5) show that sharing an institution has the highest impact on influence and knowledge.

We also regressed intellectual influence and knowledge of each referenced paper on the distinct pathways through which the author of the focal paper found the reference using ordinal logistic regressions. We use the following specification for influence and knowledge:

$$\text{logit}(P(y_{ij} \leq k)) = \alpha_i + \beta_{k0} + \beta \mathbf{X}_{ij} + \varepsilon_{ijk}$$

where $y$ represents influence or knowledge, and the vector of independent indicator variables $\mathbf{X}$ represent the various channels through which a paper was found (e.g., personal connection, a colleague, a presentation, another paper, a database search, and so on). Indices $i$ and $j$ enumerate authors and referenced papers, respectively. Author-fixed effects $\alpha_i$ denote author-specific intercepts and allow us to control for all (stable) differences between authors, including their fields. In this way, we quantify the relationship between how a referenced paper was found and the influence that paper had on the research decisions of the respondent in writing the focal paper. Coefficients are detailed in Table 3, all of which are consistent with the correlations presented below in Figure 4.

Next, we assembled a Structural Equation Model (SEM) to put these two classes of effect together. Through this structure, we simultaneously evaluate how geographical author proximity shaped the sources through which responding authors discovered referenced papers and how these sources, in turn, shaped the ultimate outcomes of knowledge about referenced papers and their influence on the focal work. The assembled SEM is estimated by the R package "lavaan" (Rosseel 2012) and has the structure indicated by Figure 3.

**Geographic Proximity**
Same Department
Same Institution
Same City
Same Country

→

**Discovery Source**
Know Personally
Colleague
Another Paper
Database
Other
Not Sure

→ Influence
→ Knowledge

Figure 4: SEM concept.

Given that all of our variables are observed, our SEM contains no latent variables. The body of individual regressions that make up the SEM are as follows:

$$\text{know_personally} \sim \alpha_{kp_i} + \text{same_department}_{ij} + \text{same_institution}_{ij} + \text{same_city}_{ij} + \text{same_country}_{ij} + \varepsilon_{kp_ij}$$

$$\text{colleague} \sim \alpha_{c_i} + \text{same_department}_{ij} + \text{same_institution}_{ij} + \text{same_city}_{ij} + \text{same_country}_{ij} + \varepsilon_{c_ij}$$

$$\text{presentation} \sim \alpha_{p_i} + \text{same_department}_{ij} + \text{same_institution}_{ij} + \text{same_city}_{ij} + \text{same_country}_{ij} + \varepsilon_{p_ij}$$

$$\text{another_paper} \sim \alpha_{ap_i} + \text{same_department}_{ij} + \text{same_institution}_{ij} + \text{same_city}_{ij} + \text{same_country}_{ij} + \varepsilon_{ap_ij}$$

$$\text{database} \sim \alpha_{d_i} + \text{same_department}_{ij} + \text{same_institution}_{ij} + \text{same_city}_{ij} + \text{same_country}_{ij} + \varepsilon_{d_ij}$$

$$\text{not_sure} \sim \alpha_{ns_i} + \text{same_department}_{ij} + \text{same_institution}_{ij} + \text{same_city}_{ij} + \text{same_country}_{ij} + \varepsilon_{ns_ij}$$

$$\text{other} \sim \alpha_{o_i} + \text{same_department}_{ij} + \text{same_institution}_{ij} + \text{same_city}_{ij} + \text{same_country}_{ij} + \varepsilon_{o_ij}$$

$$\text{influence} \sim \alpha_{inf_i} + \text{know_personally}_{ij} + \text{colleague}_{ij} + \text{presentation}_{ij} + \text{another_paper}_{ij} + \text{database}_{ij} + \text{not_sure}_{ij} + \text{other}_{ij} + \text{same_department}_{ij} + \text{same_institution}_{ij} + \text{same_city}_{ij} + \text{same_country}_{ij} + \varepsilon_{inf_ij}$$

$$\text{knowledge} \sim \alpha_{k_i} + \text{know_personally}_{ij} + \text{colleague}_{ij} + \text{presentation}_{ij} + \text{another_paper}_{ij} + \text{database}_{ij} + \text{not_sure}_{ij} + \text{other}_{ij} + \text{same_department}_{ij} + \text{same_institution}_{ij} + \text{same_city}_{ij} + \text{same_country}_{ij} + \varepsilon_{k_ij}$$

where each intercept is author-specific, and each error is author and paper-specific. This within-respondent SEM is well specified, and the estimation procedure converged quickly (180 iterations).

We use the results of this model (reported in Table 4) to select relevant variables for the reduced form fixed-effects interaction model with which we demonstrate how the interaction between sharing an institution and document similarity between focal and reference paper most significantly predicts influence and knowledge. SEM coefficient estimates, detailed below in the results, indicate that the pathway exerting the largest, most significant effect from collocation to discovery to influence and knowledge moves from sharing an institution with the author of the referenced paper, through finding the paper via personal familiarity with the author, to the influence that paper had on the focal paper and the knowledge the respondent has of its content. This confirms and adds specificity to the reduced form models detailed above and allows us to remove auxiliary, insignificant pathways from the model.

Finally, we designed a reduced form ordinal logistic regression and linear mixed models (reported in Table C1 in Appendix C) to quantify the interactive effect of institutional and intellectual distance on the respondent's (1) knowledge of the content of referenced papers, as well as (2) the extent to which the references influenced research choices reflected in the focal paper. We use the following specification for measuring influence and knowledge:

$$\text{logit}(P(y_{ij} \leq k)) = \alpha_i + \beta_{k0} + \beta_1 x_{1ij} + \beta_2 x_{2ij} + \beta_3 (x_{1ij} * x_{2ij}) + \varepsilon_{ijk}$$

where $y$ represents influence or knowledge, $x_1$ is an indicator variable for shared institution and $x_2$ represents intellectual similarity. As above, the indices $i$ and $j$ enumerate authors and referenced papers, respectively. Author fixed-effects $\alpha_i$ denote author-specific intercepts and enable us to control for all (stable) differences between authors, including their fields. This approach accounts for the possibility that the composition of authors varies significantly across the citation distribution of references. For example, different authors may have different standards for "influence." This model allows us to focus our attention, and statistical tests, on the interactions that our theory and empirical patterns of influence justify.

## 3. Results

Table 2: Regression of Influence and Knowledge on Geo-Institutional Proximity

| | *Dependent Variable:* | |
|---|---|---|
| | influence | knowledge |
| | (1) | (2) |
| *Independent Variables* | Estimate (std error) | Estimate (std error) |
| same department | 0.238 ** (.089) | 0.212 * (.086) |
| same institution | 0.764 *** (.134) | 0.810 *** (.129) |
| same city | 0.600 *** (.141) | 0.413 ** (.135) |
| same country | 0.261 ** (.087) | 0.268 ** (.084) |
| same world | 0.110 (.094) | -0.044 (.091) |
| Observations | 12309 | 12285 |
| R2 | 0.016 | 0.031 |
| Adjusted R2 | -1.466 | -1.431 |
| F Statistic | 15.652*** (df = 5; 4912) | 31.302*** (df = 5; 4896) |

*Note:* *+ p<.1, * p<.05, ** p<.01, *** p<.001*

Our first model explores the relationship between geo-organizational proximity, the likelihood that a referenced paper will have influence on a focal study, and the author's intimate knowledge with the referenced study. We find that when scientists share an organization, the chances they will report having been influenced by other scholars is maximized. Moreover, the chances that the referenced paper they are influenced by is more intellectually distant is significant. Intellectual distance is only meaningfully significant for paper pairs that share an organization. When geographical proximity between authors increases from within an organization to within a city, and to within a country, influence falls precipitously. In Figure 5 and Table 2, we present the estimated effect of a respondent fixed-effects linear regression relating discrete geographical distances with influence.

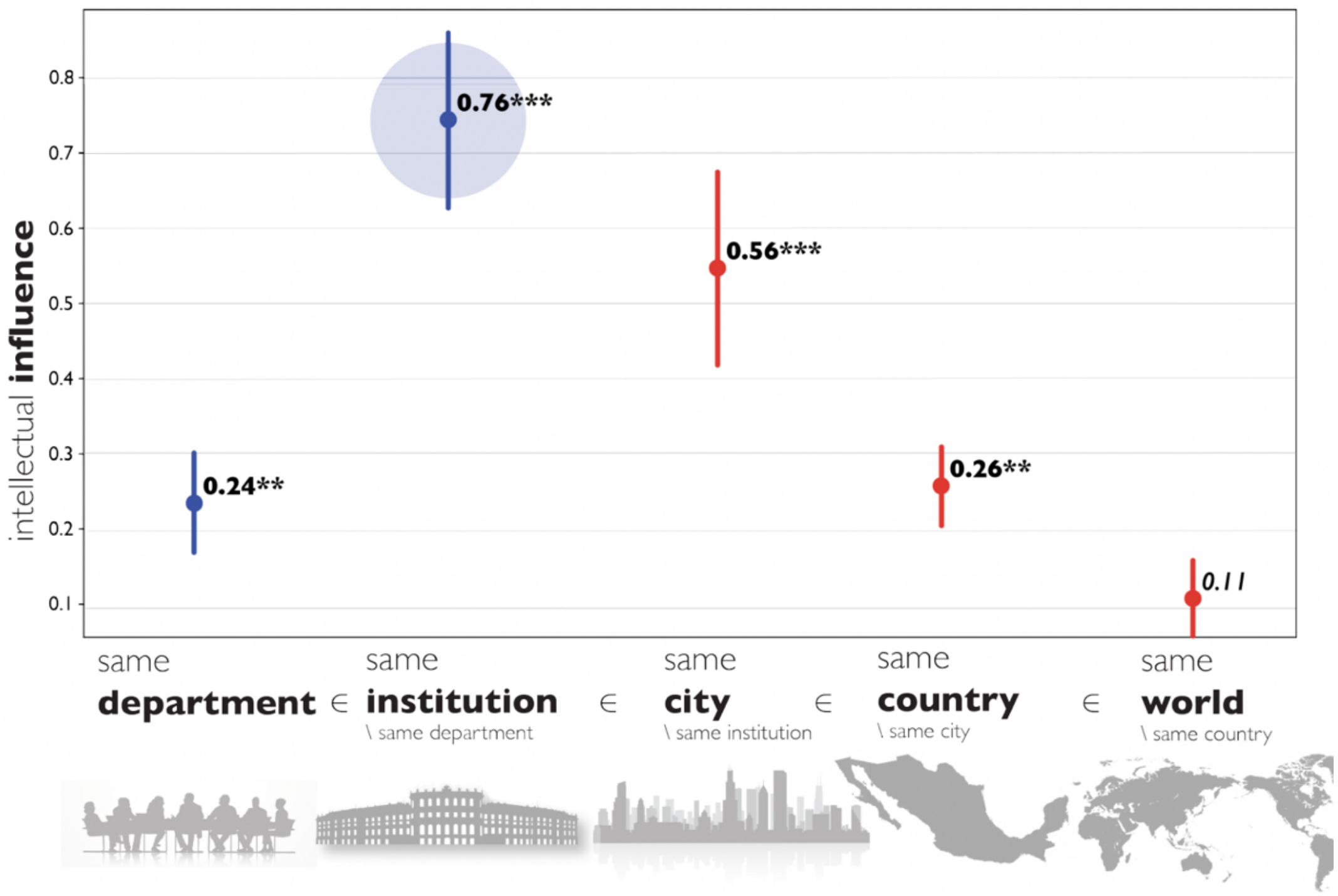

Figure 5: Effect (with std. error bars) on influence of papers at nested categorical distances from the focal paper. The first point is the effect (~5%) of intellectual influence of work cited from an author's own department(s); second the influence of work cited from an author's organizations(s) but not same department (~15%); third the influence of work cited from an author's same city but not same organization (~11%); fourth the influence of work cited from an author's same country but not same city (~5%); and finally the influence of work cited from outside the author's country (no effect). The greatest influence is from the author's organization, but not from their department or field. Being at the same organization as cited work is blue, and being at other organizations is red. Note: +p<.1, *p<.05, **p<.01, ***p<.001.

These findings suggest that institutions matter not only for access to ideas, but also for facilitating the transfer and influential absorption of those ideas. While it is significantly more likely that authors will cite works from other institutions, the works they encounter at their home institutions are the ones that influence them most. Notably, however, when authors of focal papers and reference papers share a department within the same institution, the influence they confer upon one another falls to roughly the same level as sharing a country. This is likely because colleagues within a department share enough background (Chu and Evans 2021) that their work cannot surprise and so is less likely to substantially influence.

Next, we use ordinal logistic regression models to quantify the relationship between a referenced paper's influence and an author's knowledge of it with the referenced paper's source. This model design allows us to account for the outcome variables having discrete, Likert-scale values. Results suggest that across respondents, influence and knowledge were most strongly associated with knowing the author of the referenced paper personally, learning of the paper through a presentation, and receiving it from a colleague, respectively. If we convert the coefficients into odds ratios through exponentiation, we see that knowing the author personally is associated with an increase in the odds of an additional unit of influence (on the 1-5 Likert scale) by 2.24 times and a unit of knowledge by more than 3 times. Influence and knowledge of the paper were, unsurprisingly, most negatively associated with not knowing how the respondent discovered the paper and discovering it through database search.

Table 3: Ordinal logistic regression of influence and knowledge on channels through which authors found reference papers

| | *Dependent Variable:* | |
|---|---|---|
| | influence | knowledge |
| | (1) | (2) |
| *Independent Variables* | Estimate (std error) | Estimate (std error) |
| know personally | 0.805 *** (.064) | 1.172 *** (.053) |
| colleague | 0.621 *** (.071) | 0.241 *** (.342) |
| presentation | 0.641 *** (.090) | 0.631 *** (.074) |
| another paper | 0.365 *** (.051) | 0.142 *** (.041) |
| database | 0.137 * (.054) | -0.030 (.044) |
| not sure | -0.423 *** (.114) | -0.790 *** (.093) |
| other | 0.319 *** (.073) | 0.570 *** (.076) |
| *threshold coefficients:* | | |
| 1 \| 2 | -1.401 (.060) | -1.861 (.050) |
| 2 \| 3 | 0.558 (.058) | -0.518 (.046) |
| 3 \| 4 | 2.360 (.067) | 1.088 (.047) |
| 4 \| 5 | 4.471 (.088) | 3.053 (.058) |
| Observations | 12149 | 12128 |

*Note:* + $p<.1$, * $p<.05$, ** $p<.01$, *** $p<.001$

We also allowed respondents to provide free text information on "other" ways they found referenced papers. In analyzing these self-described "other" channels, we sorted responses by the amount of influence the referenced paper had on the focal paper. Reading through responses for those that imparted a high degree of influence, we found that a much higher proportion describe personal relationships. Consider the following responses: "[M]y supervisor was the co-author [of the referenced paper]"; the author of the referenced paper is the "[f]ormer Ph.D. advisor of my co-author"; "[t]he first author [of the referenced paper] has collaborated on other projects with me." In contrast, text responses describing other ways of finding referenced papers that confer little influence tend to evince confusion and impersonal pathways. Many respondents state plainly, "I don't remember" or "I would assume that my coauthor learned about this paper via normal database search"; "searched for relevant papers, and [the referenced article] was one that came up"; or the referenced article was "suggested by a referee", signifying that, when reviewers suggest a paper, focal paper authors feel obliged to add the paper to the reference list. They do not, as a result, reimagine the project, remeasure the phenomenon, or remodel their critical outcomes.

We assembled these prior models into a structural equation model (SEM) that simultaneously estimates geo-organizational proximity and its influence on the source through which authors became aware of referenced papers. Findings from the SEM are visualized in Figure 5 and detailed in Table 4. The fitted parameters establish the link from geographical proximity, through discovery, to influence and knowledge explicitly. The model establishes reasonable relationships between proximity and discovery. Being at the same organization has the highest impact on finding a referenced paper through personal acquaintance with its author or a live presentation, and the lowest on finding it through another paper, a database, or not remembering at all. In turn, finding a paper through personal knowledge or a presentation have the highest impact on that paper's influence and an author's knowledge of it. When entire pathways are considered together, the path most responsible for influence runs from sharing an institution, through personal acquaintance with the referenced paper's author(s), to influence and knowledge.

Table 2: Structural Equation Model Results

| *Dependent Variable* | *Independent Variable* | Estimate | Std. Err | z-value |
|---|---|---|---|---|
| know_personally ~ | | | | |
| | same_department | 0.358*** | 0.037 | 9.655 |
| | same_institution | 1.580*** | 0.085 | 18.576 |
| | same_city | 1.469*** | 0.104 | 14.064 |
| | same_country | 0.555*** | 0.035 | 15.808 |
| colleague ~ | | | | |
| | same_department | 0.139*** | 0.042 | 3.329 |
| | same_institution | 0.417*** | 0.100 | 4.192 |
| | same_city | 0.450*** | 0.120 | 3.762 |
| | same_country | 0.172*** | 0.041 | 4.228 |
| presentation ~ | | | | |
| | same_department | 0.066. | 0.051 | 1.296 |
| | same_institution | 0.324*** | 0.117 | 2.766 |
| | same_city | 0.172 | 0.160 | 1.075 |
| | same_country | 0.276*** | 0.046 | 5.948 |
| another_paper ~ | | | | |
| | same_department | -0.059+ | 0.035 | -1.702 |
| | same_institution | -0.77*** | 0.125 | -6.186 |
| | same_city | -0.416*** | 0.129 | -3.229 |
| | same_country | -0.062+ | 0.034 | -1.817 |
| database ~ | | | | |
| | same_department | -0.207*** | 0.032 | -6.461 |
| | same_institution | -1.398*** | 0.099 | -14.150 |
| | same_city | -0.968*** | 0.108 | -8.944 |
| | same_country | -0.281*** | 0.031 | -8.948 |
| not_sure ~ | | | | |
| | same_department | -0.119+ | 0.061 | -1.949 |
| | same_institution | -0.49* | 0.232 | -2.115 |
| | same_city | -0.464+ | 0.277 | -1.677 |
| | same_country | -0.030 | 0.057 | 0.533 |
| other ~ | | | | |
| | same_department | 0.094* | 0.048 | 1.976 |
| | same_institution | 0.621*** | 0.100 | 6.212 |
| | same_city | 0.498*** | 0.130 | 3.843 |
| | same_country | 0.248*** | 0.044 | 5.610 |
| influence ~ | | | | |
| | know_personally | 0.186*** | 0.015 | 12.504 |
| | colleague | 0.143*** | 0.016 | 8.984 |
| | presentation | 0.189*** | 0.019 | 10.068 |
| | another_paper | 0.078*** | 0.013 | 5.944 |
| | database | -0.061*** | 0.012 | -5.081 |
| | not_sure | -0.140*** | 0.019 | -7.191 |
| | other | 0.010 | 0.017 | 0.603 |
| | same_department | -0.012 | 0.030 | -0.400 |
| | same_institution | -0.116 | 0.094 | -1.230 |
| | same_city | -0.031 | 0.105 | -0.295 |
| | same_country | -0.230*** | 0.031 | -7.522 |
| knowledge | | | | |
| | know_personally | 0.372*** | 0.015 | 24.269 |
| | colleague | 0.09*** | 0.017 | 5.363 |
| | presentation | 0.260*** | 0.021 | 12.394 |
| | another_paper | 0.048*** | 0.014 | 3.502 |
| | database | -0.109*** | 0.012 | -8.959 |
| | not_sure | -0.243*** | 0.020 | -12.057 |
| | other | 0.122*** | 0.017 | 7.342 |
| | same_department | -0.062+ | 0.033 | -1.863 |
| | same_institution | -0.241* | 0.104 | -2.324 |
| | same_city | -0.34** | 0.128 | -2.661 |
| | same_country | -0.259*** | 0.034 | -7.628 |

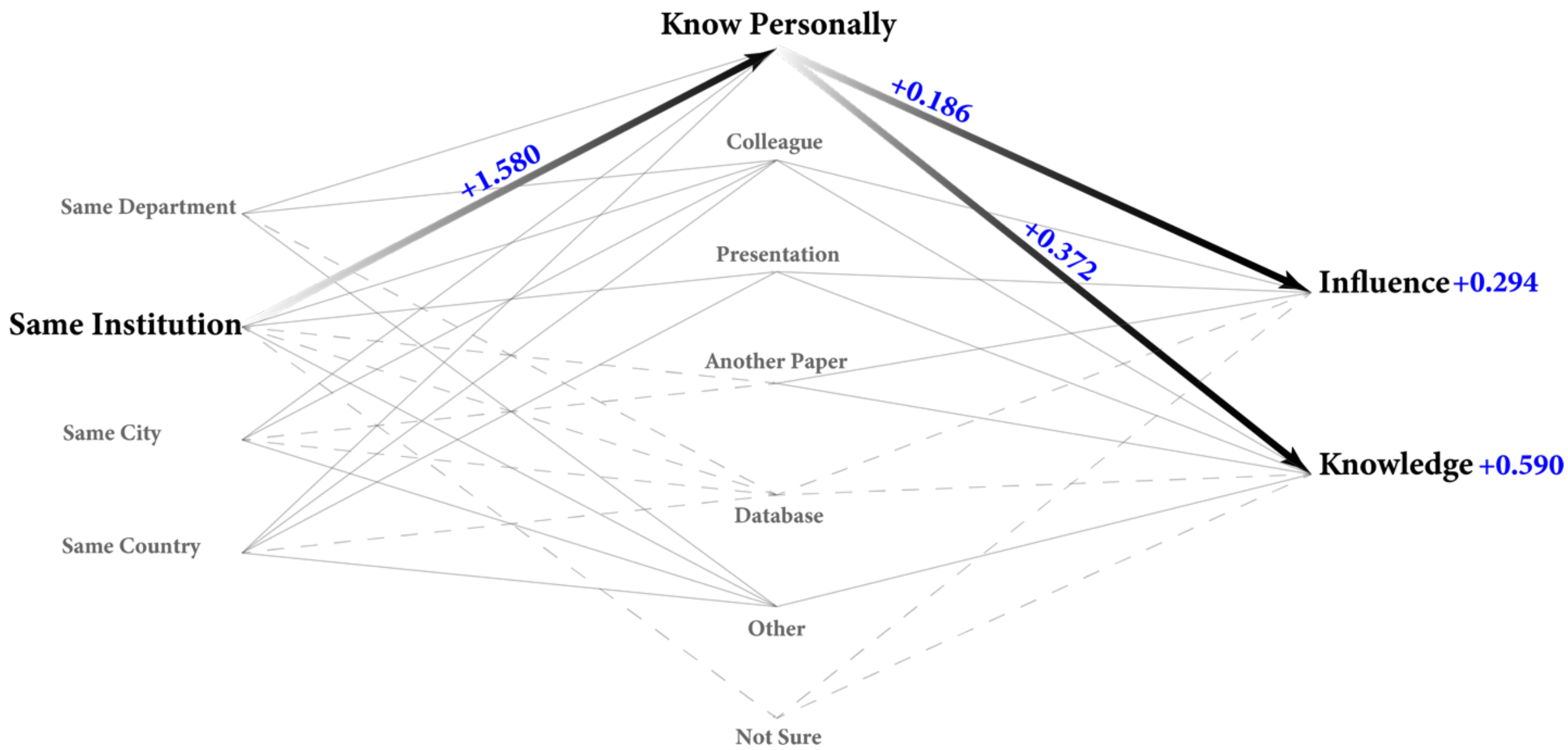

Figure 5: SEM links omitted for non-statistically significant effects. Dashed lines represent negative effects.

As a function of our SEM findings and that the only significant correlation between geo-organizational proximity and intellectual distance is shared organization, we regressed intellectual influence on selected organizational and intellectual distance measurements to identify the relationship between organizational proximity and influence. As detailed in Methods above, we perform ordinal logistic regressions to quantify the relationship between a focal paper's distances (organizational and intellectual) from its references and its influence on the corresponding author's knowledge of it and its influence on research choices, incorporating author fixed effects. We use author fixed effects for both dependent variables of Influence and Knowledge, with coefficients represented in Table 5 (see Appendix C for equivalent linear model). Both within and across individual scientists, being at the same institution with the author of research that is maximally different from your own is associated with a marked, statistically significant increase in its likelihood of influence and knowledge. The effect of sharing an institution on influence and knowledge is similarly strong.

Table 5

| | *Dependent Variable:* | |
|---|---|---|
| | influence | knowledge |
| | (1) | (2) |
| *Independent Variables* | Estimate (std error) | Estimate (std error) |
| same institution | 2.686 *** (.546) | 3.017 *** (.627) |
| document similarity | 0.337 * (.132) | 0.135 (.140) |
| same institution : document similarity | -2.132 ** (.664) | -1.577 * (.755) |
| *threshold coefficients:* | | |
| 1 \| 2 | -1.455 (.114) | -2.479 (.125) |
| 2 \| 3 | 0.458 (.111) | -0.814 (.119) |
| 3 \| 4 | 2.207 (.114) | 1.232 (.119) |
| 4 \| 5 | 4.271 (.126) | 3.512 (.127) |
| Observations | 12309 | 12285 |
| *Note:* | *+ p<.1, * p<.05, ** p<.01, *** p<.001* | |

Table 5: Regression results from ordinal logistic regression on sharing and institution, increasing document similarity, and the interaction effect (indicated by ':') of sharing an institution and increasing document similarity on influence and knowledge.

If we convert the coefficients into odds ratios through exponentiation, we see that being at the same institution is associated with an increase in the odds of an additional unit of influence (on the 1-5 Likert scale) by 14.7 times and a unit of knowledge by 20.4 times. The effect of document similarity (in isolation) on influence varies between the ordinal and linear models (slightly increasing influence by 1.4 times in the ordinal model; and slightly decreasing it in the linear model), which does not affect the claims of this paper. Being at the same institution and having more document similarity strongly decreases the likelihood of influence by a factor of 8.4 and knowledge by a factor of 4.8, suggesting that a likely path to influence and knowledge involves being at the same institution and sharing less document similarity.

These results explain the observed increase in influence within institutions when moving from pairs of papers that share an academic department to pairs that do not (Figure 4). Universities and research institutes matter because they connect people beyond departments, whose work is intellectually distant from one another. Institutions facilitate influentially distant and often chance intellectual encounters— such as showing up at an unexpected presentation on campus—underlying a disproportionate share of scientific and scholarly advance.

## 4. Discussion

In summary, we find that for all fields, when focal paper authors share an institution with the author of a reference paper, the more intellectually distant the reference paper, the more influence that paper has on the focal paper. Moreover, scientists and scholars are significantly more likely to find those intellectually distant papers at their home institutions. This provides strong evidence that institutions like universities and research institutions play an outsized role in the catalysis of cross-disciplinary knowledge discovery and influence.

Our study has natural limitations. For one, we use self-reports of intellectual influence and familiarity. Nevertheless, the design of the sampling and solicitation process reduces self-selection and reporting biases: we explicitly compare randomly sampled citations from the same paper, the respondent was not free to select the paper(s) or citations they report on, and we confirm that they cited and remembered cited papers, adding a layer of explicit verification atop self-reported citations. This improves upon the established survey approach of asking respondents to identify an instance of a phenomenon in question (e.g., a case of discrimination, influence, etc.) and then answer questions relevant to that case, which otherwise leads to a focus on extreme or subjectively salient instances. By contrast, we randomly sample from the space of acknowledged influences, validate their recognition of that cited influence, and ask details about it relative to another sampled, cited influence from the same source. An obvious limitation of this sampling strategy is that it cannot select papers respondents read but did not cite. While the degree to which each researcher is susceptible to influence and how they interpret the survey questions may be different, we control for this by asking respondents about two referenced papers and perform regressions with respondent fixed-effects where possible. These "within-author" models ensure that observed differences are not confounded by endogenous citing tendencies or idiosyncratic definitions of "influence." Finally, in this paper, we focus on the general effect of sharing an institution with researchers with diverse intellectual backgrounds. While this effect is observed across the 15 fields represented in our study, an evaluation and exploration of field level variation is beyond the scope of this paper, and represent both promising and interesting areas for exploration in future work.

Recent scholarship suggests that allowing for geographic flexibility via remote work leads to increases in output and employee satisfaction (Choudhury, Foroughi, and Larson 2021; Bloom et al. 2014; Möhring et al. 2021; Barrero, Bloom, and Davis 2021), but this misses the importance of influence and knowledge transfer for innovative advance as observed in (Yang et al. 2021). Our investigation demonstrates why. Sharing an institution is a critically important meso-scale for intellectual exposure and influence between the micro-scale of sharing an office, hallway, or department and the macro-scale of sharing a city, state, or country. While (van der Wouden and Youn 2023) observe that co-location increases learning in collaboration, we observe that co-location increases both learning and influence regardless of whether co-located individuals collaborate.

The meso-level of organizational collocation matters more than any other we examine for facilitating the transfer of influence in science. Organizations matter by promoting occasions for interaction between diverse intellectual viewpoints through committees, seminars, gyms, and dining halls—the work of the university and the often insular communities that serve them (Owen-Smith 2018). At the micro-scale of the office next door and the macro-scale of the international scientific congress, researchers interact with others more intellectually similar to themselves. The value of critical meso-scales has been observed in online communities like Wikipedia, where the institutional constraint of a single article for a single topic—the work of producing an encyclopedia—necessitates interaction between diverse viewpoints, which is in turn associated with higher quality encyclopedia articles (Shi et al. 2019). At the micro- and macro-scales of the internet as a whole, we see the opposite, with ideological echo chambers serving as the paradigm of what can happen when proper institutional constraints for the promotion of sustained, diverse interactions are not in place (Bishop 2009; Sunstein 2001; Bail et al. 2018). Our work builds on and theorizes prior work by (Rawlings and McFarland 2011; Kabo et al. 2015b) that finds that researchers are more likely to be productive when they learn and adopt best practices from their peers, and that shared interactions are more likely to lead to proposed innovations and how universities can encourage this by promoting research collaboration. Finally, our work both extends and clarifies the findings of (Wuestman, Hoekman, and Frenken 2019; Leahey 2006). Here, we discover that the optimal distance for influence, while a form of co-location, is at the organizational, not the departmental level.

In particular, our work suggests that when researchers learn from their most intellectually distant peers, such pedagogical encounters are the most efficacious for unleashing flows of intellectual influence.

This result helps to not only theorize but provide an antidote to the observed tension between the opportunity cost of being geographically distant and the dampening effect on innovation from being too geographically close (Esposito and Rigby 2018). As a result, geographical proximity associated with universities (Owen-Smith 2018) is a core ingredient in producing not only influence, but sustainable innovation by exposing scientists to intellectually distant ideas. The importance of this ingredient can easily be missed with more blunt instruments such as citation counts, but becomes clear when we focus directly on what such instruments attempt to capture—influence.

Being physically proximate to others that do very different—apparently unrelated—research at one's own university dramatically increases the likelihood that their work comes to influence and potentially drive one's published discoveries. In recent years, the importance of place has been enshrined within built infrastructure for interdisciplinary engagement (Mäkinen, Evans, and McFarland 2020), but comparable investments have been made in inter-institutional research networks and distributed "centers" of excellence around the world.

Our findings alongside recent scholarship contradict recent commentary in the popular press, as in the *New York Times*, where an investigative piece asked the question "Do Chance Meetings at the Office Boost Innovation?" and answered, "There's No Evidence of It." The piece quoted, at length, a scholar of workplace interaction and transparency who stated, "there's credibility behind the argument that if you put people in spaces where they are likely to collide with one another, they are likely to have a conversation, but is that conversation likely to be helpful for innovation, creativity, useful at all for what an organization hopes people would talk about? There, there is almost no data whatsoever. All of this suggests to me that the idea of random serendipity being productive is more fairy tale than reality" (Miller 2021). Our findings provide data that suggest the innovative power of in-place encounters for science. They demonstrate that sharing a space leads researchers to discover papers by those with whom they share it, that these papers are more likely to be intellectually distant, and that they, in turn, confer greater influence and knowledge.

The results presented in this study document the value of fostering sustained diversity in place. In this age of continuing COVID-19 care and ongoing debate over the importance of being physically together, if we hope to continue to fuel the engine of innovation, we will need to replace, and not simply displace, this essential but underappreciated mechanism of influence operating within our physical universities. We need to find new ways to consistently engage with potential influences that may seem unrelated, but could be crucial for sustaining our continued research and understanding of the natural and social world.

## Appendix A. Survey materials

The following text and figures illustrate the survey flow. Figure A1 displays the (anonymized) recruitment email.

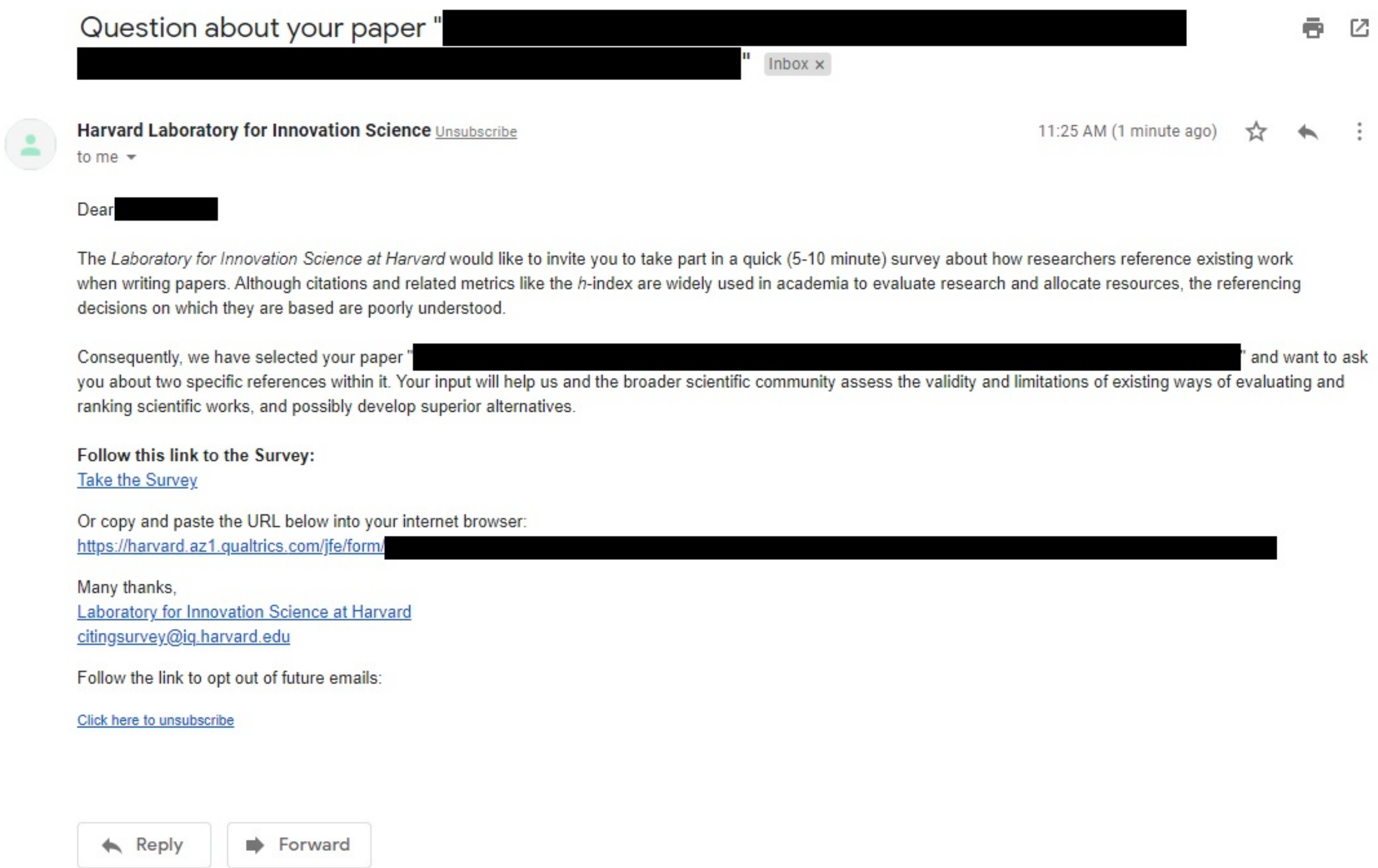

Question about your paper "[redacted]" Inbox ×

**Harvard Laboratory for Innovation Science** Unsubscribe
to me
11:25 AM (1 minute ago)

Dear [redacted]

The *Laboratory for Innovation Science at Harvard* would like to invite you to take part in a quick (5-10 minute) survey about how researchers reference existing work when writing papers. Although citations and related metrics like the *h*-index are widely used in academia to evaluate research and allocate resources, the referencing decisions on which they are based are poorly understood.

Consequently, we have selected your paper "[redacted]" and want to ask you about two specific references within it. Your input will help us and the broader scientific community assess the validity and limitations of existing ways of evaluating and ranking scientific works, and possibly develop superior alternatives.

**Follow this link to the Survey:**
Take the Survey

Or copy and paste the URL below into your internet browser:
https://harvard.az1.qualtrics.com/jfe/form/[redacted]

Many thanks,
Laboratory for Innovation Science at Harvard
citingsurvey@iq.harvard.edu

Follow the link to opt out of future emails:

Click here to unsubscribe

Reply Forward

Figure A1: Sample recruitment email.

After clicking on the link, respondents proceeded to confirm that the paper was indeed theirs and read IRB information. Next, they proceeded to a randomized page, the two versions of which are displayed in Figure A2. The control (panel A) and treatment (panel B) versions are identical except that treatment includes the reference's citation information.

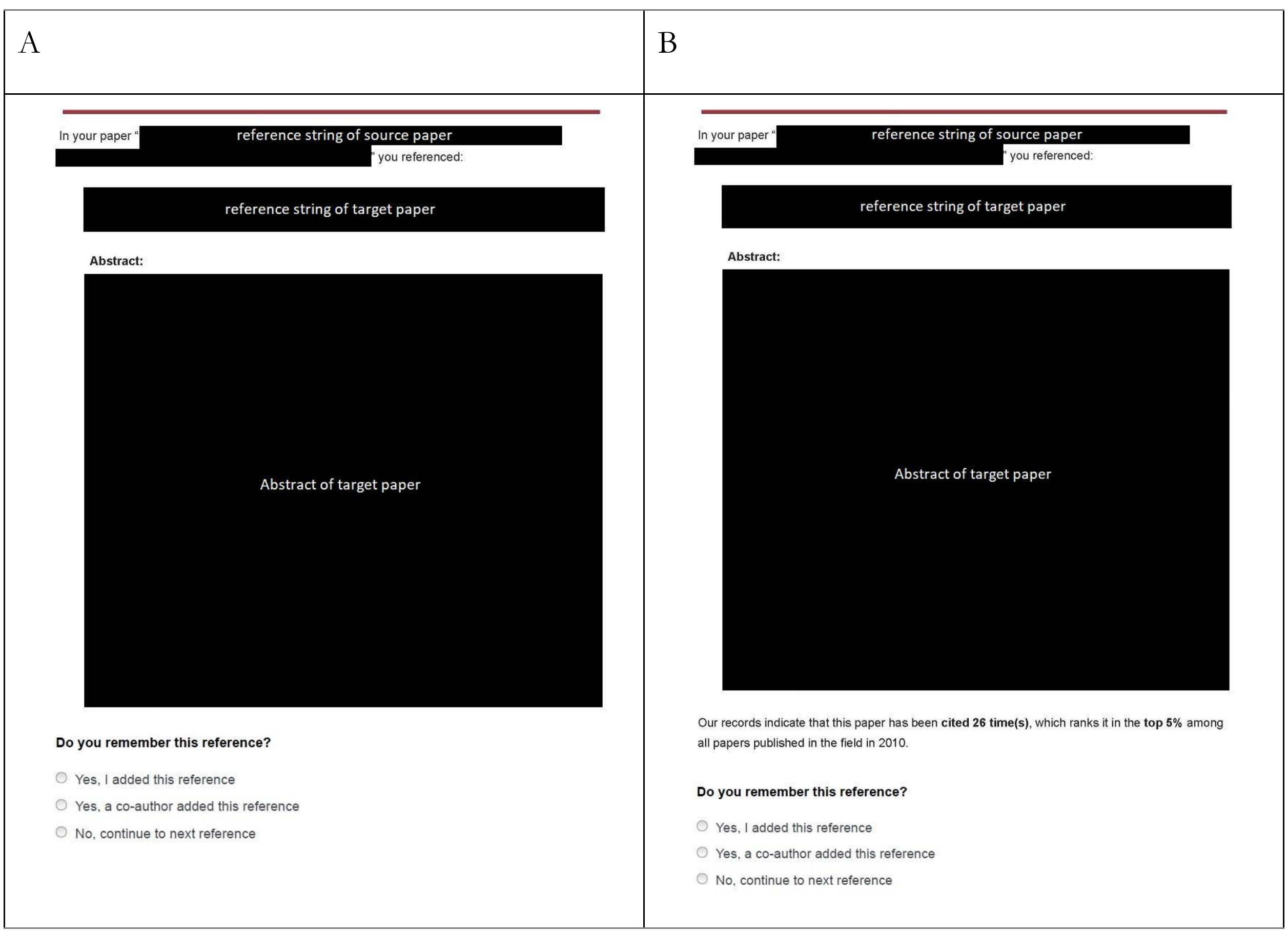

Figure A2: Two forms used for the reach signal experiment. 85% of randomly assigned respondents saw the control form (Panel A), which does not show any citation information, and 15% saw the treatment form (Panel B), which displays the true citation count and percentile.

Next, respondents answered questions about their knowledge of the reference, how much it influenced them, which aspects of their work were influenced (Figure A3). To account for ordering effects in answer choices, respondents were randomized into two forms with identical questions but reversed answer choice order. Form A's answer choices ranged from smallest/least to biggest/most, while form B had the opposite ordering. Next, respondents rated the reference on various dimensions of quality (Figure A4), described their expertise in the reference and how/when they first discovered it (Figure A5). Lastly, respondents provided some demographic information.

Reference:

reference string

**How well do you know this paper?**

- **Extremely well** (know it as well as my own work)
- **Very well** (familiar with all findings, data & methods, all limitations and critiques)
- **Well** (familiar with all findings, data & methods, some limitations)
- **Slightly well** (familiar with all findings, data & methods)
- **Not well** (only familiar with main findings)

**How much did this reference influence the research choices in your paper?**

- **Very major influence** (motivated the entire project)
- **Major influence** (influenced a core part of paper, e.g. choice of theory or method)
- **Moderate influence** (influenced an important part of the paper, e.g. additional analysis)
- **Minor influence** (influenced a small part of paper, e.g. added sentence(s) to Discussion)
- **Very minor influence** (paper would've been very similar without this reference)
- **Not sure**

**Which aspects of your paper did this reference influence?** (Mark all that apply)

- Only minor influence
- Topic
- Theory or conceptualization
- Data
- Methods
- Other. Please explain

Figure A3: Screenshot illustrating questions about the author's knowledge of the reference and its impact on the author. Randomly assigned half of the respondents saw this ordering of answer choices, while another half saw the reverse ordering.

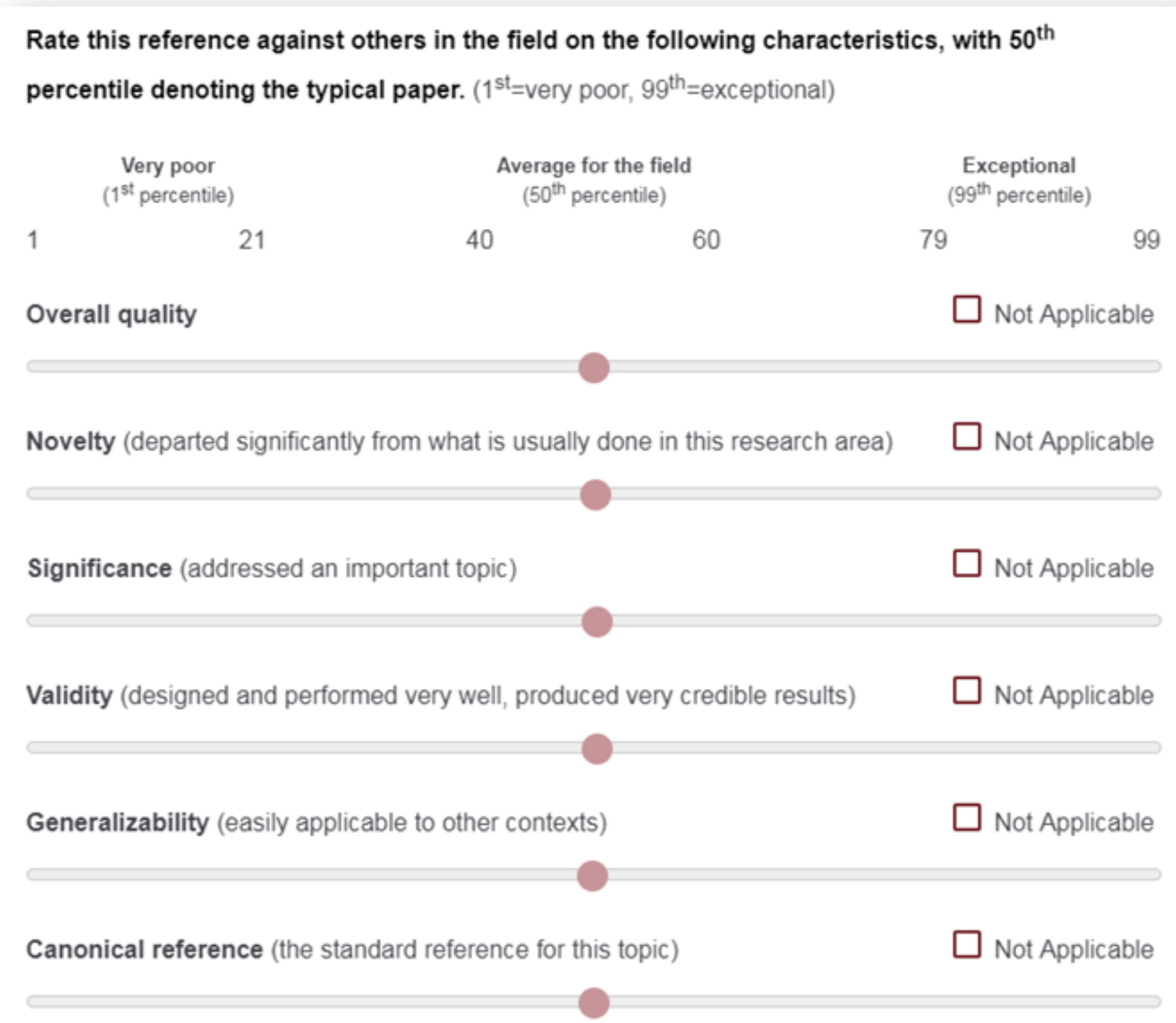

Figure A4: Panel of questions about perceived quality of the reference. The attribute in the last position was randomized to be “Canonical” or “Prominent.” Data from this last position is not included in the present analyses due to its indirect relationship with quality, but is available from the authors upon request.

**How much expertise do you have in the topic(s) of this reference?**

- Very high expertise (inside my core research area)
- High expertise
- Moderate expertise (at the boundary of my research area)
- Some expertise
- Very little expertise (outside my research area)
- Not sure

---

**How did you (or the co-author who added this reference) first learn about this reference?**
(Mark all that apply)

- Self-citation
- Know the author personally
- Recommended by a colleague
- Database search (e.g. Google Scholar)
- Saw in a list of references of another paper
- Saw in a conference, presentation, or class
- Not sure
- Other. Please explain

**When did you (or the co-author who added this reference) first learn about this reference?**

- Before the project started
- During the project's *early* stages (e.g. design, data collection)
- During the project's *middle* stages (e.g. analysis)
- During the project's *late* stages (e.g. drafting the manuscript)
- During review/publication process
- Not sure

Figure A5: Questions about respondent's expertise in the topic(s) of the reference, and how and when the respondent first learned about the reference.

## Appendix B. Response Analysis

Disciplines

Response rates were measured by clicks on the personalized survey link. Rates varied substantially across disciplines. The lowest response rate came from oncology (12.9%) and the highest (34.1%) from history and philosophy of science. The number of completed responses and response rates by discipline are displayed in Figure B1.

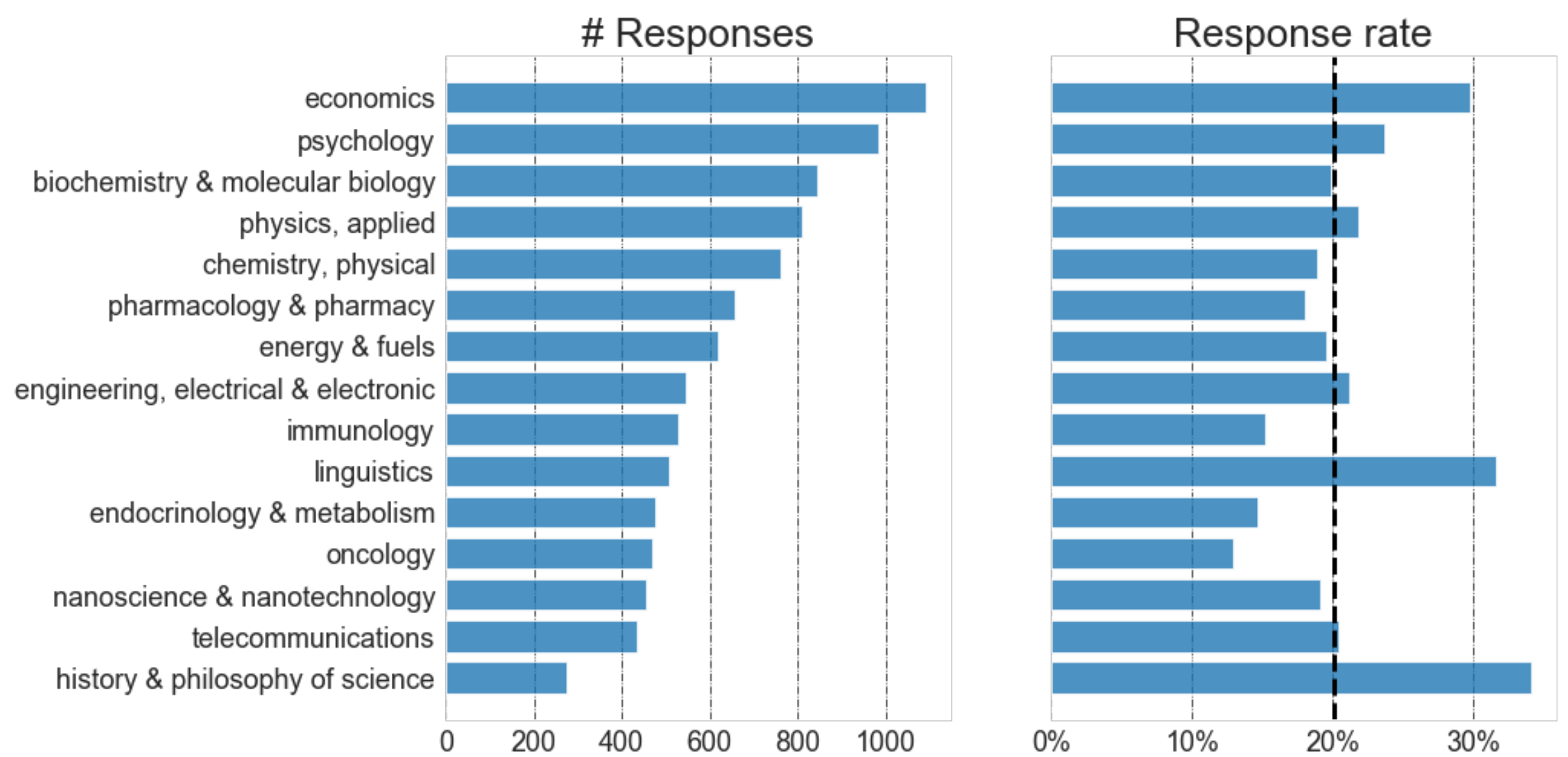

Figure B1: Response counts and response rates by discipline. Each response, if filled out completely, provides data on two references. The dotted line shows the mean response rate.

Table B1

| Discipline | Responses on Influence & Knowledge |
|---|---|
| Economics | 1078 |
| Psychology | 1096 |
| Biochemistry & Molecular Biology | 1060 |
| Applied Physics | 864 |
| Physical Chemistry | 1361 |
| Pharmacology & Pharmacy | 834 |
| Energy & Fuels | 1419 |
| Engineering, Electrical & Electronic | 688 |
| Immunology | 622 |
| Linguistics | 421 |
| Endocrinology & Metabolism | 589 |
| Oncology | 701 |
| Nanoscience & Nanotechnology | 497 |
| Telecommunications | 569 |

| History and Philosophy of Science | 209 |
|---|---|

Table B1: Responses that included data on 'Influence' and 'Knowledge'. Degree of survey completion varied with some respondents failing to answer certain questions. Responses to questions concerned with Influence and Knowledge represented in this table are included in the results reported in the paper.

For self-selection and non-response analysis, see (Teplitskiy et al. 2022) from which response data used in this study are derived.

## Appendix C. Influence, Knowledge & Source

Respondents were asked how they found the papers that they surveyed them about. The options are represented in Figure C1, below.

**How did you (or the co-author who added this reference) first learn about this reference?**
(Mark all that apply)

- ☐ Self-citation
- ☐ Know the author personally
- ☐ Recommended by a colleague
- ☐ Database search (e.g. Google Scholar)
- ☐ Saw in a list of references of another paper
- ☐ Saw in a conference, presentation, or class
- ☐ Not sure
- ☐ Other. Please explain

Figure C1: Survey question recording responses on how respondents first learned of the references.

In addition to the ordinal logistic regression presented in Table 5 in the main body of the article, here we model the effect and interaction of sharing an institution and increasing document similarity on influence and knowledge using linear panel data models with respondent specific fixed-effects:

$$influence_{ij} = \alpha_i + \beta_0 \text{ same institution } + \beta_1 \text{ document similarity } + \beta_3 \text{ (same institution } * \text{ document similarity) } + \epsilon_{ij}$$

$$knowledge_{ij} = \alpha_i + \beta_0 \text{ same institution } + \beta_1 \text{ document similarity } + \beta_3 \text{ (same institution } * \text{ document similarity) } + \epsilon_{ij}$$

We observe that both within and across individual scientists, being at the same institution with the author of research that is maximally different from your own is associated with an increase in its likelihood of influence by roughly 63% —33% linked to being at the same institution, 8% linked to being dissimilar in content, and 27% linked to both being at the same institution *and* having dissimilar content. Moreover, being at the same institution is associated with an increase in knowledge of cited content by nearly 19% while reading dissimilar content is linked to increasing knowledge of that content by roughly 7%.

Table C1

| | *Dependent Variable:* | |
|---|---|---|
| | influence | knowledge |
| | (1) | (2) |
| *Independent Variables* | Estimate (std error) | Estimate (std error) |
| same institution | 1.654 *** (.334) | 0.949 *** (.323) |
| document similarity | -0.422 *** (.155) | -0.329 ** (.140) |
| same institution : document similarity | -1.372 ** (.420) | -0.207 (.407) |
| Observations | 12309 | 12285 |
| R2 | 0.012 | 0.013 |
| Adjusted R2 | -1.475 | -1.474 |
| F Statistic | 19.597*** (df = 3; 4914) | 22.242*** (df = 3; 4898) |

*Note:* + $p<.1$, * $p<.05$, ** $p<.01$, *** $p<.001$

Table C1: Effect and interaction effect of sharing an institution and increasing document similarity on influence and knowledge.

Table C2: Correlation Table

| | influence | knowledge | geographic distance | intellectual distance | same dept | know personally | colleague | presentation | another paper | database | not sure | other | same institution | same city | same country | same world |
|---|---|---|---|---|---|---|---|---|---|---|---|---|---|---|---|---|
| influence | 1.0*** | | | | | | | | | | | | | | | |
| knowledge | 0.544*** | 1.0*** | | | | | | | | | | | | | | |
| geographic distance | -0.002 | -0.062*** | 1.0*** | | | | | | | | | | | | | |
| intellectual distance | 0.032*** | 0.003 | -0.025* | 1.0*** | | | | | | | | | | | | |
| same dept | 0.038*** | 0.053*** | -0.037*** | -0.011 | 1.0*** | | | | | | | | | | | |
| know personally | 0.125*** | 0.235*** | -0.203*** | 0.006 | 0.083*** | 1.0*** | | | | | | | | | | |
| colleague | 0.085*** | 0.056*** | -0.058*** | -0.001 | 0.03 | 0.057*** | 1.0*** | | | | | | | | | |
| presentation | 0.09*** | 0.115*** | -0.048*** | 0.003 | 0.013 | 0.157*** | 0.044*** | 1.0*** | | | | | | | | |
| another paper | 0.05*** | 0.024** | 0.036*** | 0.028** | -0.013 | -0.056*** | -0.013 | 0.047*** | 1.0*** | | | | | | | |
| database | -0.058*** | -0.1*** | 0.127*** | -0.019* | -0.055*** | -0.309*** | -0.235*** | -0.142*** | -0.209*** | 1.0*** | | | | | | |
| not sure | -0.058*** | -0.105*** | 0.003 | -0.001 | -0.016+ | -0.072*** | -0.059*** | -0.041*** | -0.079*** | -0.221*** | 1.0*** | | | | | |
| other | 0.01 | 0.072*** | -0.062*** | -0.02* | 0.017+ | 0.004 | -0.053*** | -0.037*** | -0.106*** | -0.263*** | -0.032*** | 1.0*** | | | | |
| same institution | 0.053*** | 0.1*** | -0.161*** | -0.042*** | -0.059*** | 0.193*** | 0.037*** | 0.025** | -0.054*** | -0.136*** | -0.018* | 0.06*** | 1.0*** | | | |
| same city | 0.047*** | 0.055*** | -0.148*** | 0.009 | 0.016+ | 0.142*** | 0.034*** | 0.007 | -0.027** | -0.08*** | -0.015+ | 0.034*** | -0.015+ | 1.0*** | | |
| same country | -0.013 | 0.028** | -0.457*** | -0.007 | 0.047*** | 0.14*** | 0.038*** | 0.056*** | -0.016+ | -0.079*** | 0.005 | 0.05*** | -0.005 | -0.049*** | 1.0*** | |
| same world | -0.034*** | -0.085*** | 0.401*** | 0.02* | -0.638*** | -0.218*** | -0.059*** | -0.05*** | 0.042*** | 0.139*** | 0.016+ | -0.067*** | -0.208*** | -0.167*** | -0.652*** | 1.0*** |

*Note:* + $p<.1$, * $p<.05$, ** $p<.01$, *** $p<.001$

Table C2: Complete correlation table of all variables used in this study.